\documentclass{amsbo}
\usepackage[dvips]{graphicx}
\usepackage{mathptmx}
\usepackage{amsmath}
\usepackage{amssymb}

\newcommand{\um}[1]{\ensuremath{\,\mathrm{#1}}} 

\title{Search for Fractional Charges in Cosmic Rays with Ams}

\author{Cristina Sbarra\footnote{\texttt{Cristina.Sbarra@bo.infn.it}}, \\
   D.Casadei, L.Brocco,  A.Contin, G.Levi, F.Palmonari}

\affil{INFN - Sezione di  Bologna, Viale Berti Pichat 6/2,
I-40126 Bologna, Italy}

\shorttitle{Contributed Paper presented at the 18th European Cosmic Rays Symposium, Moscow - 2002}
\shortauthors{C.~Sbarra}
\date{April, 2002}
\renewcommand{\NoteNumber}{Presented at the 18th ECRS, Moscow - 2002}

\begin{abstract}

  Preliminary results on the flux of non strongly-interacting,
  fractionally charged particles in primary cosmic rays at 400 Km
  above sea level are given.  Cosmic ray data collected by AMS-01 in June
  1998 have been analysed on the hypotheses of 2/3 charged
  leptons.  The search is carried on by looking at the energy
  deposition measurements by the time of flight system scintillator
  counters. A preliminary flux limit is given.  

\end{abstract}


\begin{document}

\maketitle

\tableofcontents

\section{Introduction}
Many historical experiments stated that the charge of matter comes in
discrete units, and determined the amount of those `quanta' of charge
(by Millikan \citep{1}). Later on, several reasearches have been
made in order to test if, in some circumstances, a fractional amount of
that elementary charge could be produced.

From the theoretical point of view, in the Standard model of quantum
chromodinamics, there is space only for colour singlet particlese.
Althout the quarks, which are the bricks of the model, come with
fractional charge, up to now they have been seen to be confined into
mesons and barions integrally charged.  Few grand unified theories,
anyway, account for color singlet particles with fractional charge
(see \citep{12}, \citep{13} and \citep{14}).  Moreover, some theories
of spontaneously broken QCD have predicted free quarks \citep{17},
although in the standard quark interactions scheme, those free quarks
should soon strongly interact with matter, and would hardly penetrate
a particle detector.

Leptons known so far, are only neutral or with charge multiple of e.
Anyhow, a fractional lepton could produce ionization in a detector,
penetrating a large amount of material, leaving an energy proportional
to $Q^2$ and, in some cases, passing the detection efficiency of the
apparatus.  So, it is worthwhile searching for lightly ionizing
particle among the whole data taken by a particle detector although it
was born with another aim.  Any observation of fractional charge, in
fact, would be a direct evidence of physics beyond the standard model.

\section{The AMS-01 time of flight (TOF)}
In June 1998, the Alpha Magnetic Spectrometer, in its first version
(AMS-01), was carried for ten days on board of the shuttle space mission
(STS91), collecting an amount of 100 millions of cosmic rays.

\begin{figure}
\begin{center}
  \includegraphics[width=1.1\columnwidth]{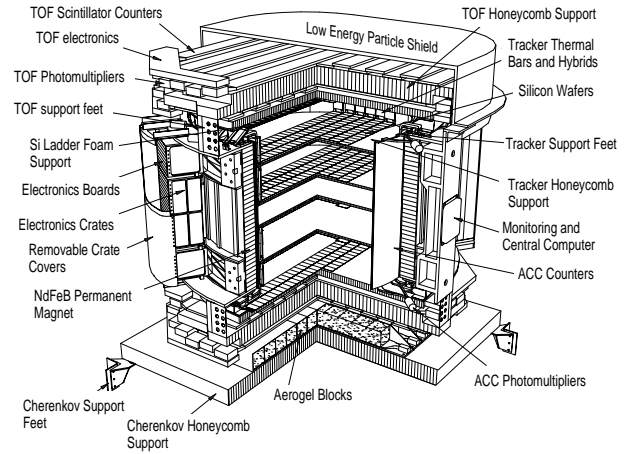}
\caption{The AMS detector for the STS-91 mission (AMS-01).}
\label{f:AMS-01}
\end{center}
\end{figure}

The detector of AMS-01, whose configuration is shown in figure \ref{f:AMS-01}, 
consists of: a permanent magnet equipped with six layers of silicon
tracker, that measures the trajectory of relativistic particles with
an accuracy of 10 \micron\ in the bending direction and 30 \micron\ in
the non-bending one; a scintillator system for the rejection of events
due to interation in the magnet inner walls; a threshold Areogel
Cerenkov system, and a Time Of Flight (TOF) system.

The TOF consists of four planes which measure the transit time of
charged particles with a resolution of 120 \um{ps} over a distance of
1.4 \um{m}.  The TOF also yelds multiple layers energy loss
measurements, providing the fast trigger for the AMS experiment.  Four
TOF scintillators layers and up to eight silicon tracker layers, in
fact, measure $\frac{dE}{dx}$, allowing a multiple determination of
the absolute value of the particle charge.

The AMS trigger logic can be divided in these steps: first the
scintillator data are processed in a very fast way, and perform what
is called the \emph{Fast Trigger} (FT), that gives the `time zero' of
the experiment. The \emph{first level} trigger, that togheter with the
presence of the FT also asks for the absence of the anticoincidence
data.  Then, the \emph{third level} that precesses the data from the
various detectors: asks for AND of same counter sides in the TOF,
applies a cut on the maximum curvature of the trajectory in the
tracker (this request leaves about 14\% of the total triggers),
controls that the track from tracker fits with the track from TOF.

The Fast trigger is thus made only by the TOF, when al least one side
of one counter, in each plane, has gained the ``High Threshold'' (HT)
of 150 mV.  The Ams data are recorded when also other trigger levels
are gained (except for the so called 'prescaled events', about 1/100
of the total, where only the fast triggers was requested). Usually,
the requirement for the trigger were three over four planes of the
TOF. In correspondence of a trigger, however, all of the four planes
were always recorded, thus allowing the search for ``rare events'' in
the whole data taken during the STS91 flight so far.

\begin{figure}
\begin{center}
  \includegraphics[width=1.1\columnwidth]{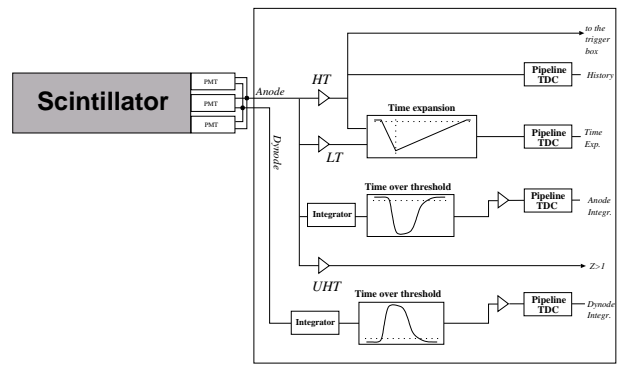}
\caption{Logic of the TOF counters readout board.}
\label{f:TOF1readout}
\end{center}
\end{figure}

\section{Hypothesis for the analisys}
In order to give a limit on the flux of rare events, it has been made
the hypothesis that the fractionary charges (2/3 e) have an enegy loss
distribution which is a 'properly scaled' proton landau. Moreover, it
has been used tha same Accptance as for the AMS-01 proton analisys
(and so the cuts). For this analisys, finally, only the Fast Trigger
effiuciency changes and has to be computed.

A cosmic ray flux is given by:
\begin{equation}
\label{flux}
\phi = \frac{N_{obs}}{\epsilon \cdot  A \cdot T}
\end{equation}

where $A$ is the Acceptance, $T$ is the observation time, $N_{obs}$ is
the events observed.  thus, being:

\begin{equation*}
\epsilon=\epsilon_{FT}\cdot\epsilon_{LVL3}\cdot \epsilon_{cut}
\end{equation*}

And, for our assumptions, being
\begin{equation*}
A\cdot\epsilon_{LVL3}\cdot\epsilon_{cut}\simeq 0.15 \; m^2 sr 
\end{equation*}

then, in our analisys, it had to be computed only the fast trigger
efficiency and the observation time.

\subsection{Tipical energy loss in TOF planes}
\begin{figure}
\begin{center}
  \includegraphics[width=1.1\columnwidth]{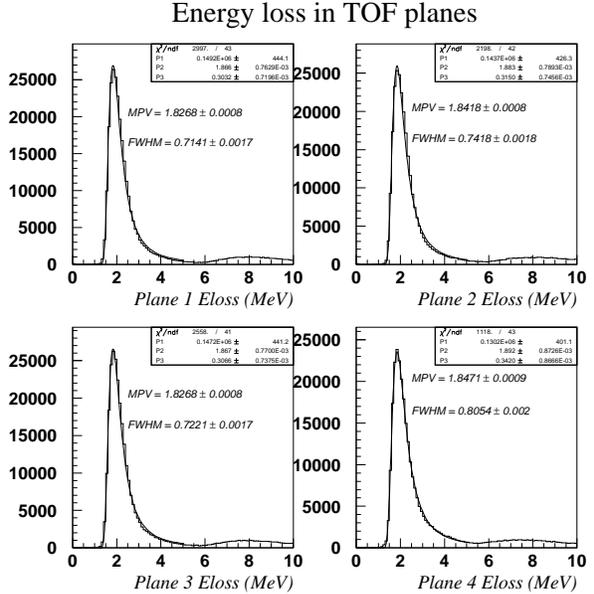}
\caption{Tipical energy loss of triggered particles traversing the TOF planes.}
\label{landmev}
\end{center}
\end{figure}

The tipical energy loss of triggered particles traversing the TOF
planes is given in figure \ref{landmev}. As you can see, the Most
Probable Value (MPV) of 1 MIP, in each plane, is about 1.8 MeV. From
now on, we can pass to MIP units, in the whole analisys.

\subsection{Fast trigger efficiency}
\begin{figure}
\begin{center}
  \includegraphics[width=1.1\columnwidth]{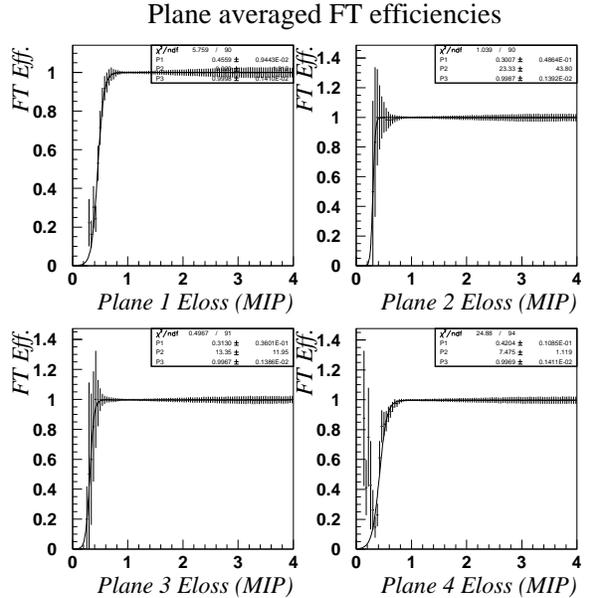}
\caption{Fast Trigger Efficiency of the four TOF planes.}
\label{effmip}
\end{center}
\end{figure}

It is possible to compute the FT efficiency for detecting particles
with low charge depositions in each of the four planes of the TOF. To
do this, the method of the ``spectator plane'' is used: everytime a
plane is not in the 3/4 triggered planes, it is possible to look at
its recorded data to see if the HT was passed or not. In this way at
the end we have the FT efficiency of the plane, in MIP units, shown in
figure \ref{effmip}.

Later on, the total FT efficiency is
the sum over the combinations of three planes, formally:

\begin{equation*}
\begin{aligned}
\label{eq:toteff}
\epsilon  = \binom{3}{4}\cdot \biggr( \epsilon^{pl1}\cdot \epsilon^{pl2}\cdot \epsilon^{pl3}+ 
 \epsilon^{pl2}\cdot \epsilon^{pl3}\cdot \epsilon^{pl4}+  \\
\epsilon^{pl1}\cdot \epsilon^{pl2}\cdot \epsilon^{pl4}+ 
 \epsilon^{pl1}\cdot \epsilon^{pl3}\cdot \epsilon^{pl4} \biggr) 
\end{aligned}
\end{equation*}

\section{Expected energy loss distribution for $\frac{2}{3} e$}

considering the energy loss distribution of $\frac{2}{3}$ e charges,
as a ``scaled proton landau'', and taking into account the efficiency
of each plane, we can get the expected energy loss distribution of
those fractional charges, in each plane (see, for plane 1, figure
\ref{laneffpl1}). From that figure it is evident that, if we put a cut
at 0.7 MIP, we can be confident to be away from the proton background.
Then, dividing the integral of the expected flux that is on the left
of the cut, by the total integral of the ``scaled'' proton landau, we
get the FT efficiency for $\frac{2}{3} e$ for that plane.

\section{Lightly ionizing particle in the TOF}

It has been plotted the energy released in the TOF planes, after some
canonical cuts (the same used for the AMS-01 proton analisys) shown in
figure \ref{cuts}, looking at the low energy losses.  The ``canonical
cuts'' consist basically in asking for data taken away from the south
athlantic anomaly, with at least 1 cluster per TOF plane, and with
only 1 TOF bar involved in each cluster. Later on, only few tracker
cuts are added (``track quality'' cuts), but they just reduce the
events of a 10\%. The plots are shown in figure \ref{loweloss}, with a
mark at 0.7 MIP.

\begin{figure}
\begin{center}
  \includegraphics[width=1.1\columnwidth]{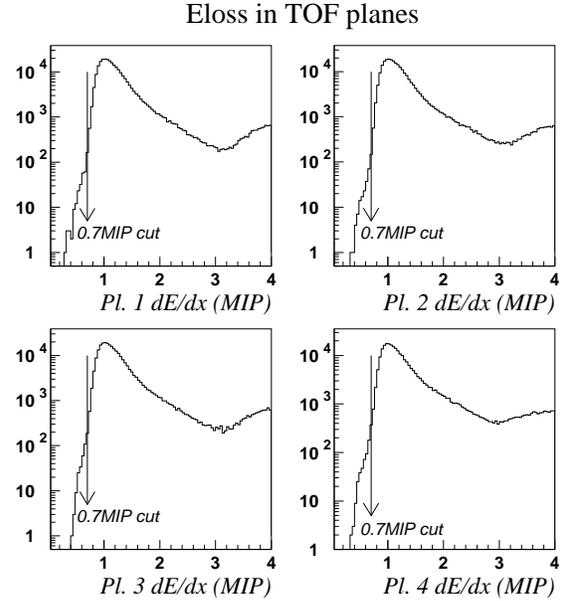}
\caption{Energy released in TOF (in MIP units) with mark at 0.7 MIP}
\label{loweloss}
\end{center}
\end{figure}

\begin{figure}
\begin{center}
  \includegraphics[width=1.1\columnwidth]{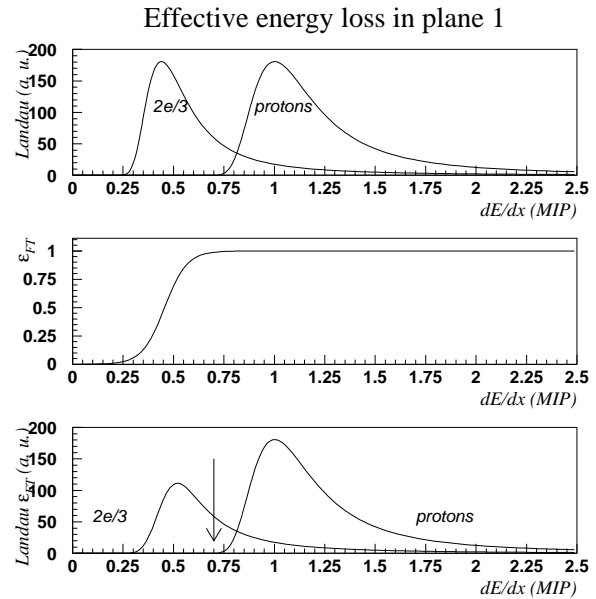}
\caption{Expected Energy loss distribution in plane 1 of the  TOF, for $\frac{2}{3} e$ and for protons (in MIP units). The TOF efficiency FT of plane 1, is multiplied for the scaled landau.}
\label{laneffpl1}
\end{center}
\end{figure}

We now have to compute the fast trigger efficiency for detecting
$\frac{2}{3} e$. It is possible to think on the expected energy loss
of $\frac{2}{3}$ as a properly scaled proton landau, as tou can see in
figure \ref{laneffpl1}.

Here are the various plane efficiencies calculated in the mentined way:
\begin{equation*}
\begin{aligned}
\epsilon^{pl1}=0.43 \\
\epsilon^{pl2}=0.71 \\
\epsilon^{pl3}=0.68\\
\epsilon^{pl4}=0.47
\end{aligned}
\end{equation*}
that led to the combined efficiency of FT, for detecting fractionally
charge: $\epsilon _{{FT},{\frac{2}{3}}}=0.18$

\section{Analysis and results}

The AMS-01 data have been analized using basically the same cuts that
have been used in the proton AMS-01 official analisys.  They consist
on data taken away from the south atlantic anomaly, no overlaps of the
TOF counters, and finally only few 'track quality cuts' of traker,
that reduce of a 10\% only the events.  At this last point has been
evaluated the low energy loss in the TOF planes (plotted in figure
\ref{loweloss}). As you can observe from the plot, below tha 0.7 MIP
there are still some candidates in the four TOF planes. Nevertheless,
if we require a low energy loss simulaneously in the first three TOF
planes, then the bulk of candidates disappears.

In such cases of counting ``rare events'', where a poisson variable
$n$ counts signal events with unknown mean $s$ (and with $\theta$
prior) as well as background with mean $b$, then, in the absence of a
clear discovery (e.g., when $n$=0 or if $n$ is compatible with the
expected background), it is possible to give an upper limit on $s$. In
substance, when $n=0$, the mean $s$ is, at 95 \% of C.L., less than 3.
This results from even bayesian approach than frequentists one
(\citep{dagostini}, \citep{pdg2001}).

We finally calculated the observation time, which in this case is
$T=36436 \; s$,

\begin{figure}
\begin{center}
  \includegraphics[width=1.2\columnwidth]{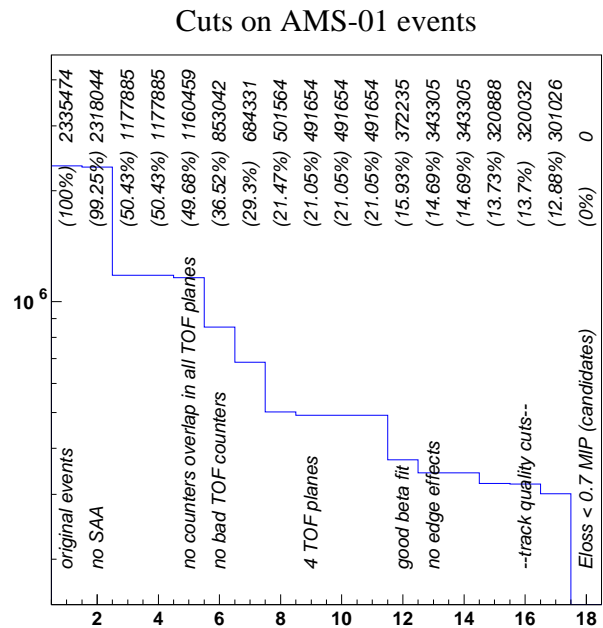}
\caption{Energy released in TOF (in MIP units) with mark at 0.7 MIP, the analisys is made on the first 12 hours of AMS-01 data.}
\label{loweloss}
\end{center}
\end{figure}

\section{Conclusions}

Looking at the energy lost in the four planes of the Time of Flight of
the AMS-01 experiment (on the test flight of 1998) and selecting the
events triggered only by the TOF (the 'prescaled events'), it has been possible to put an upper limit on the $\frac{2}{3} e$
lepton-like particles flux is:
\begin{equation*}
\phi < 3.0 x 10^{-7} \; cm^{-2} s^{-1} sr^{-1}
\end{equation*}

at 95\% of Confidence Level.

This upper limit could be lowered extending the analysis on all the AMS-01 data (about factor of 10). Moreover, with the AMS-02 experiment on the International Space Station which will last for three years, the same limit can improve another factor of 100.


\begin{thebibliography}{1}
\bibitem{1}
R. Millikan, Philos. Mag. \textbf{19}, 209 (1910)
\bibitem{12}
P.H.Frampton and T.Kephart, Phys. Rev. Lett. \textbf{49}, 1310 (1982)
\bibitem{13}
S. M. Barr, D.B. Reiss, and A. Zee,Phys. Rev. Lett. \textbf{50}, 317 (1983) 
\bibitem{14}
H. W. Yu, Phys. Lett. \textbf{142B}, 42 (1984)
\bibitem{17}
A. De Rujiula, R. Giles, and R. Jaffe, Phys. Rev. D \textbf{17}, 285 (1978)

\bibitem{amsfirst}
S.P.   Ahlen \emph{et al.},  Nuclear Instruments and Methods \textbf{A
350} (1994) 351.

\bibitem{dagostini}
G.D'Agostini, Yellow Report Cern 99-03
.

\bibitem{pdg2001} 
PDG 2001, avaiable on the PDG WWW pages (URL: http://pdg.lbl.gov/)

\end{thebibliography}
\end{document}